# Exploring Symmetry in Wireless Propagation Channels

Ehab Salahat[1], Ahmed Kulaib[1,2], Nazar Ali[2], Raed Shubair[2,3]

[1] Etisalat-British Telecom Innovation Center, Abu Dhabi, UAE
[2] Electrical and Computer Engineering Department, Khalifa University, UAE
[3] Research Laboratory of Electronics, Massachusetts Institute of Technology, USA

*Abstract*—Wireless communications literature is very rich with empirical studies and measurement campaigns that study the nature of the wireless propagation channel. However, despite their undoubted usefulness, many of these studies have omitted a fundamental yet key feature of the physical signal propagation, that is, wireless propagation *asymmetry*. This feature does not agree with the *electromagnetic reciprocity theorem*, and the many research papers that adopt wireless channel symmetry, and hence rendering their modeling, unexpectedly, inaccurate. Besides, asymmetry is unquestionably an important characteristic of wireless channels, which needs to be accurately characterized for vehicular/mobile communications, 5G networks, and associated applications such as indoor/outdoor localization. This paper presents a modest and a preliminary study that reports potential causes of propagation asymmetry. Measurements conducted on Khalifa University campus in UAE show that *wireless channels* are *symmetric* in *the absence* of symmetry impairments. Therefore, care should be taken when considering some practical wireless propagation scenarios. Key conclusions and recommendation are summarized. We believe that this study will be inspiring for the academic community and will trigger further investigations within wireless propagation assumptions.

## I. Introduction

Signal propagation in wireless channels can be subjected to many types of environmental parameters that degrade its performance. Such factors include noise, interference, large-scale fading (shadowing), small-scale fading, path loss, delay, and other temporal and spatial dynamics of the link that act as impairments to the propagated electromagnetic signal [1]. As such, modeling radio channels is unquestionably a challenging endeavor due to the combined effect of these impairments, and an accurate characterization will be of a great benefit for the design of resilient future 5G wireless and vehicular networks, protocols, and applications that will work in practice.

While experimental results that are based on hardware implementation and empirical measurements clearly achieve the best realism, practical considerations such as ease of use, controllability, fine-tuning, repeatability and configurability have made simulation software the dominant evaluation tool, especially in dynamic networks [1]. Therefore, the design of these networks depends on mathematical and conceptual models that can predict their performance in real environments [2]. However, the lack of sophisticated simulation models, negligence of proper statistical analysis, use of unjustified and overly simplified assumptions as well as the adoption of some theorems of wireless communications have led to uncertainties in these studies that are buttressed with graphs and results produced by these unrealistic simulation models. In fact, the broad chasm between simulation and reality calls into question the validity of many of the (seminal) research papers, and the applications that work well based on these simulation models are highly unlikely to work in practice [3] [4] [5].

One of the common assumptions that is typically adopted in the wireless communications technical literature is channel symmetry. Specifically, between two transceiving nodes, say $\mathcal{A}$ and $\mathcal{B}$, some of the propagation impairments are assumed to have an equivalent effect (or are just ignored) when the roles of the transmitting and the receiving ends are swapped (while keeping the same link setup in terms of the transmission frequency, transmitter and receiver gain, and measurement distance) resulting in identical channel responses. Such an assumption also follows Friis free-space transmission (a.k.a. link budget) model [6]:

$$P_r(d) = P_t \mathcal{G}_t \mathcal{G}_r \left[\frac{\lambda}{4\pi d}\right]^2, \quad (1)$$

where $P_t$, $\mathcal{G}_t$, $\mathcal{G}_r$, and $d$ respectively denote the transmitter power, transmitter gain, receiver gain, and the distance. This model is incorporated into the path loss model in any given measurement environment, which is analytically given as [6]:

$$P_r(d, \beta) = P_r(d_0)\left(\frac{d_0}{d}\right)^\beta, \quad (2)$$

where $\beta$ and $d_0$ denote the path loss exponent (e.g. $\beta = 2$ in free-space) and the reference distance, respectively. The assumption of channel symmetry also coincides with the well-known electromagnetic reciprocity theorem [7] that states, "*if the role of the transmitter and the receiver are switched, the instantaneous signal transfer function between the two remains unchanged.*"

However, the literature is very rich with counter examples that clearly does not agree with this theorem, which show that most of the wireless communication channels (and especially low-power ones) are typically *asymmetric* (e.g. [2][8][9][10]). Many counter examples that provide suitable evidence for channel asymmetry can, for instance, be taken from the Wireless Sensor Networks (WSN) [2][11][12] as well as Wireless Mesh Networks (WMN) [13]. The effect of asymmetry was also considered in many reported performance evaluation studies in the literature (e.g. [3][8][14].

Radio asymmetry can have a big effect on many sensitive applications, e.g. frequency synchronization [15][16][17] and indoor/outdoor localization (which is very critical for location-based services in vehicular and mobile networks, Internet-of-Things) using time-of-arrival (ToA) and time-difference of arrival (TDoA), where both the forward and reverse channel response affect the accuracy of these sensitive algorithms. To the best of the authors' knowledge, all literature on localization algorithms has assumed channel symmetry.

Therefore, it is important to clarify the confusion of the wireless channels asymmetry especially with respect to the reciprocity theorem and some of the reported empirical studies. To the best of our knowledge, the literature also lacks a systematic and detailed research that addresses the impact of radio asymmetry in wireless networks. The aim of this research, which extends the results reported in [18], is to clarify these issues and explain the potential and most dominant causes of channel asymmetry, which may help in establishing new wireless applications and algorithms. The contributions of the paper include:

1) presenting a brief survey on the potential causes of channel asymmetry which will be studied later in this paper,

2) highlighting such not so accurate belief by giving a simple illustrative example based on localization, which is important to vehicular and future 5G networks and applications,

3) providing outlines of a recommended, well-controlled measurements setup that can be used to conduct any further investigations. The measurements dataset itself and testing codes are publicly available at [19] which will help interested researchers to verify the results reported herein,

4) empirically showing based on measurements carried out by the authors the correctness of the electromagnetic reciprocity (that is, the *electromagnetic propagation*, when symmetry impairments are *eliminated,* is *symmetric*). Yet, it is shown that wireless propagation channels are asymmetric in nature as realistic propagation always has such symmetry impairments,

5) and finally, a modest recommendation is made to guide in developing realistic simulation models and tools while avoiding any wireless communications mistaken axioms such as the symmetry of wireless propagation channels.

The rest of this paper is structured as follows. Section II presents a detailed review of the possible channel asymmetry causes (henceforth, termed as symmetry impairments) as well as a short case study on the effects of channel asymmetry on localization. Section III includes well-controlled empirical studies in different LOS and NLOS indoor locations (which can provide pointers for accurate modeling) to minimize or eliminate the effect of these impairments. In section IV, some of the results obtained from measurements carried out at the university campus are illustrated and discussed. Finally, section V concludes the paper.

## II. Channel Asymmetry – A Synopsis

In this section, a summary of the potential causes of the asymmetric channel behavior is briefly discussed below. Then, a demonstrative localization example is provided where the knowledge of the asymmetry is critical for proper system modeling and algorithmic design.

### A. Symmetry Impairments

The wireless communications literature, as mentioned earlier, is rich with examples that show asymmetric channels or links. For example, the authors in [14] use this asymmetry term to denote unequal signal strength (or signal-to-noise ratio, SNR) and different fading conditions for the different links. In this work, and in order to avoid ambiguity, the term "*symmetry*" defines whether the instantaneous characteristics of the forward and reverse channels due to the propagation impairments (interference, multipath effects, etc.) in a specific medium are the same or not.

*i. Transmit Power*

Asymmetric transmit power (e.g. a mobile station with limited transmit power versus a base station with virtually infinite transmit power) can cause disparity in the received signal. Wireless networks are composed of a heterogeneous set of devices, and asymmetric links are very likely due to this form of asymmetric transmit powers in those devices [20].

*ii. Hardware Sensitivity*

The hardware units used in the measurements itself can be a possible cause of asymmetry in wireless networks. A study in [20] showed that even homogeneous devices (i.e. devices that are manufactured by the same vendors with the same model) but sold at different times differ in their transmission behavior. Hardware sensitivity was reported to be a major contributor to the asymmetry in [2]. An example of a hardware component that is likely to introduce asymmetry is the low noise amplifiers (LNA) that largely determines the receiver's noise floor [20].

*iii. Aperiodic Measurements*

Aperiodic measurements (i.e. taken with a long separation) is a possible cause of asymmetry in the links. For example, a mobile station with fully charged power will typically perform better than that with a battery that is about to die. Aperiodic measurements lack accuracy and hence can't be considered reliable as it will probably increase the link's asymmetry [13].

*iv. Antenna Design and Configuration.*

Another source of link asymmetry could be the used transmit and/or receive antenna [20], especially with multiple antenna systems (i.e. spatial diversity such as in SIMO, MISO, and MIMO), even though asymmetry could also be observed with SISO systems. Specifically, the degree of asymmetry can depend on the directivity of the antenna and its configuration, and the utilized algorithms to estimate channel conditions. In addition, antenna's height and orientation are potential causes of channel asymmetry, as was reported by Stuber [21]. On the other contrary, the authors in [2] stated that antenna is not responsible for an appreciable contribution of the asymmetry, based on their measurements.

*v. Interference and Noise*

The two potential contributors to link asymmetry are noise and interference [3][11][20]. Interference and noise are likely to cause temporal and site-specific variations and typically hard-hit (due to their bursty nature) the received propagated signal, unlike the previous impairments. Moreover, every radio and communications engineer knows that you erect your antenna on top of the tallest nearby location (e.g. a hill) so it has direct connectivity with all other nearby radios, or in short, high "fan in" [3]. A base station transmitting a signal from that hill will not be subjected to interference similar to that of a mobile station transmitting in a congested and crowded city. For example, if there is an interfering signal in the vicinity of a device $\mathcal{A}$, then signals from a remote device $\mathcal{B}$ to $\mathcal{A}$ might be disrupted, whereas signals from node $\mathcal{A}$ to $\mathcal{B}$ are normally strong enough to overcome the interference, and this scenario itself can be reversed overtime [13]. On the other hand, many receivers utilize LNAs, which are designed to amplify the incoming signal without compromising much on the noise. However, since LNAs are not perfect, they also amplify any time-varying artifacts in the received signal, which may introduce some level of asymmetry [20].

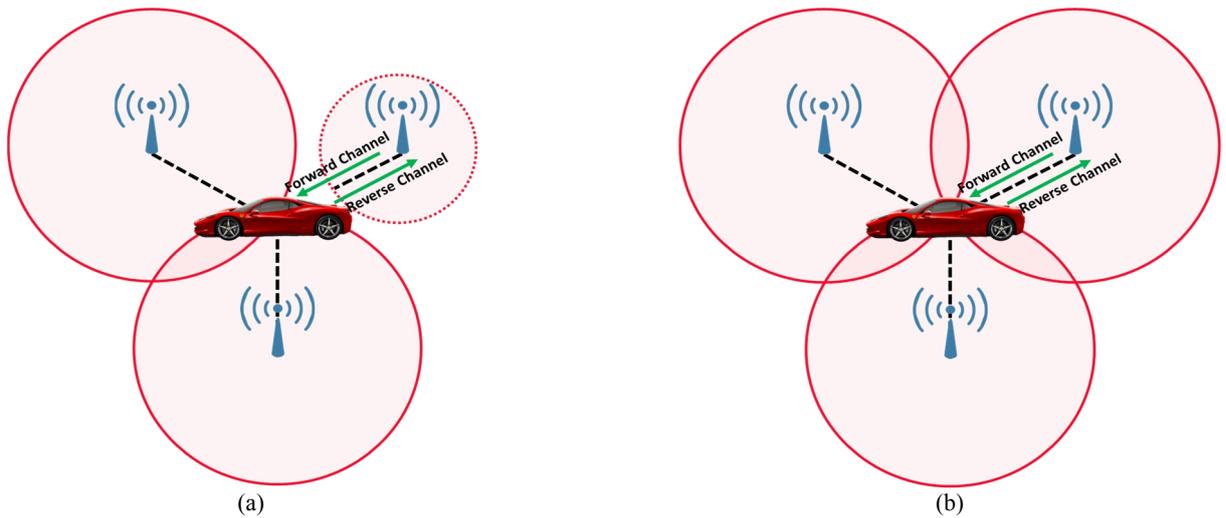

Fig. 1: Example of (a) failed localization due to incorrect time estimation based on the forward and reverse channels symmetry assumption, and (b) a more accurate estimate based on the asymmetry channels.

*vi. Spectrum Shifting and Frequency Mismatch*

A study [11] showed that "a large variation is manifested not only in the transmitted power, but also in the operational frequencies" of their WSN. This mismatch (misalignment) may be correlated with the hardware or the transmit power.

*B. An Illustrative Example*

To elucidate the severity of channel asymmetry, consider the following example adopted from the localization literature. One of the well-known ways to locate a mobile station (MS) such as a vehicle, is via triangulation, where it should fall into the range (radio vicinity) of at least three base stations (BS), as illustrated in Fig. 1. Each BS tries to estimate the distance from the MS using two-way ranging. Specifically, using the well-known fundamental physics, one knows that the speed (of an electromagnetic wave, which is $3\times10^8$ m/s) equals to the distance traveled between the BS and MS (needed for the triangulation) divided by the total travel time. The travel time is estimated from the power delay profile [6], which is obtained from the inverse Fourier transform of the channel frequency response. Since the overall travel time is between the BS to the MS plus vice versa, then the channel responses of the BS to the MS (forward channel) and vice versa (reverse channel) are needed. If we adopt the channels symmetry assumption, then travel time from BS to MS will be the same of that from the MS to BS, which is incorrect, as shown in Fig. 1 (a). However, if the asymmetric channels model is adopted, then those travel times will likely to be different (which is more realistic), and hence any erroneous distance measure can be avoided or minimized and triangulation (and hence localization) is less likely to fail, as in Fig. 1 (b).

## III. MEASUREMENT SETUP

A similar setup to the one described in [22][23] has been used to conduct frequency channel measurements. Based on the earlier discussion on the potential symmetry impairments, measurements in the frequency domain were conducted at Khalifa University building in Sharjah, UAE using a Vector Network Analyzer (VNA) from Rohde & Schwarz. Two identical omnidirectional antennas from National Instruments were positioned about 1.5 meters above the floor and connected to VNA using very low insertion loss cables. The VNA was setup to measure 10 consecutive sweeps each containing 601 frequency sampling points to obtain detailed enough data for measurements for the forward and reverse channel responses ($S_{21}$). Moreover, since the VNA is acting as a transceiver, any issue that is related to hardware sensitivity, transmit power or spectrum shifting will affect both the forward and the reverse channels equally, and so the hardware impairments are no longer an issue. In addition, the antennas used are omnidirectional with excellent performance and were mounted on approximately the same height (so the directivity and antenna orientation are longer an issue for this study). The effect of the wires that were used to connect the antennas to the VNA was eliminated via hardware calibration.

Lastly, to eliminate the effect of the interference, VNA was used to scan the WiFi frequency spectrum from 2.4~2.5 GHz (no signal was transmitted, and so any received signal was an interfering signal). The spectrum looked as shown in Fig. 2, which is obtained by observing the aforementioned frequency range over a long period of time. Given this spectrum, one can notice that the last range of frequencies, and specifically in the 2.48~2.5 GHz band, is interference free. Hence, to rule out the effect of the interference, this sub-bandwidth was used herein for the analysis. Note also that any other interference-free bandwidth can be used.

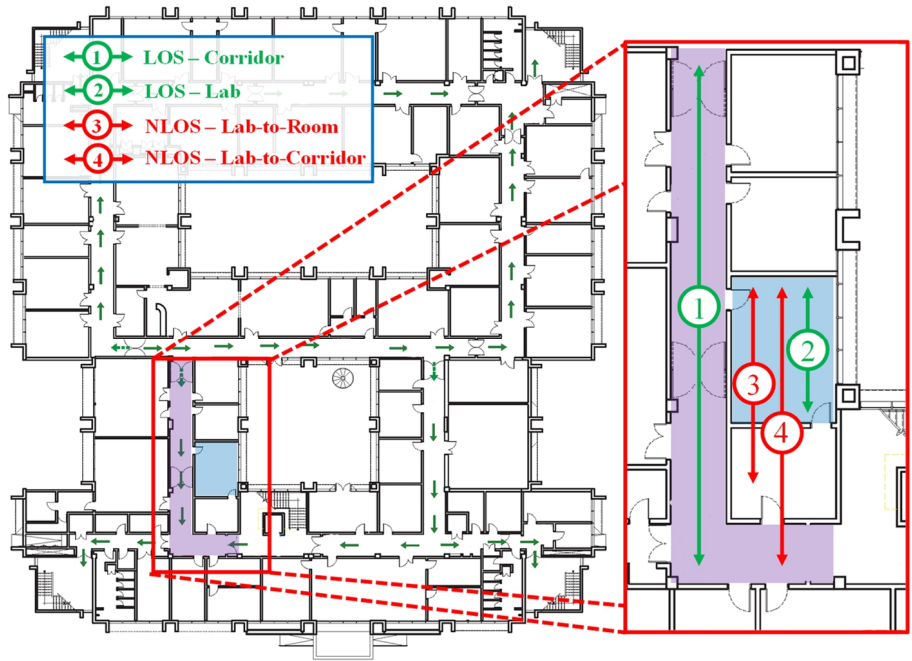

Fig. 3: Floorplan of Khalifa University (Sharjah Campus), and the indoor measurement locations for both the LOS and NLOS scenarios.

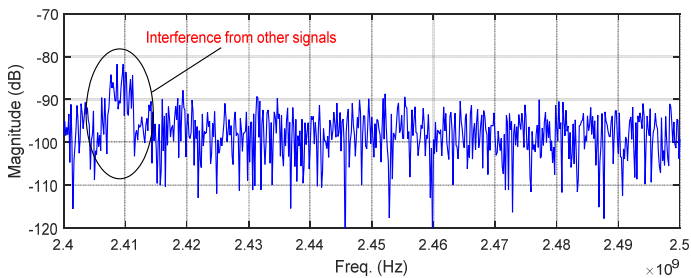

Fig. 2: Scanned WiFi spectrum (2.4~2.5 GHz).

## IV. MEASUREMENTS: RESULTS AND ANALYSIS

Given the experimental setup discussed earlier in the previous section, multiple indoor measurements at Khalifa University were conducted from which we report a few due to space limitations (however, the entire set of measurements and testing codes are available online and are accessible in [19]). Specifically, two test scenarios considered LOS environments (a corridor and an electronics lab, given in purple and blue respectively in Fig. 3), and the remaining two considered NLOS (from the lab to the adjacent room, and from the lab to the corridor through the adjacent room, in other words, one and two walls (representing light and heavy NLOS signal propagation, respectively). Exemplary results are shown in Fig. 4 to Fig. 13. Given that most of the well-known symmetry impairments were eliminated, one shall expect to see a symmetric forward and reverse channels, which can be observed in Fig. 4 – Fig. 13 as well as the other measurements skipped in this section. This does not only apply to the LOS scenarios but also to the (light and heavy) NLOS propagation scenarios regardless of the distance separating them. The symmetry can be even observed on the pathloss plots in Fig. 12 for the scenario 2 (LOS, inside the lab) and Fig. 13 for scenarios 3 and 4 (from the lab through the walls). The results given from these measurements data draw important conclusions. Specifically, the measurements:

1) verified the electromagnetic reciprocity theorem, as both the forward and reverse channels showed identical responses,

2) suggest that the multipath fading effect is not a contributor to channel asymmetry, as one may have thought, because it existed in our environments and basically cannot be eliminated in real scenarios,

3) showed that the separation and the LOS and NLOS setup has no effect on the channel asymmetry, as the forward and reverse channels showed no difference in their responses regardless of the transmitter-receiver separation,

4) even though the experiment environments were static (no motion), dynamic environments are likely not to cause any asymmetry, as the measurements are done very fast (in $\mu$s), and hence the dynamic environment is virtually static,

5) and more importantly, in realistic wireless scenarios and mobile networks, which are rich with symmetry impairments that cannot be removed as was done in our well-controlled experiments and measurements (e.g. operators have no control on the users' hardware of their smart phones), the forward and reverse wireless channels are expected to be asymmetric, due to the existence of these impairments, which corroborate the results of the cited literature (e.g. [3]).

## V. CONCLUSIONS AND FUTURE WORK

In this paper, measurements were used to prove that wireless channels can only be symmetric in the absence of common temporal and spatial impairments. It was shown, with the aid of measurements data from well-controlled experimental setup, that if the effect of these impairments is eliminated (which is not the case in realistic scenarios), then the electromagnetic propagation will be symmetric, agreeing with the reciprocity theorem. The paper has also shown that the multipath fading, transmitter-receiver separation, and LOS/NLOS propagation

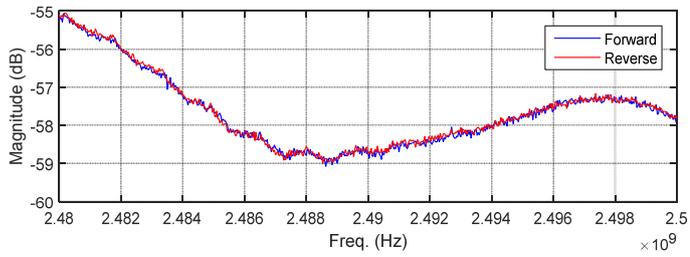
Fig. 4: Scenario 1 in Fig. 3, measurements, separation distance=4 m.

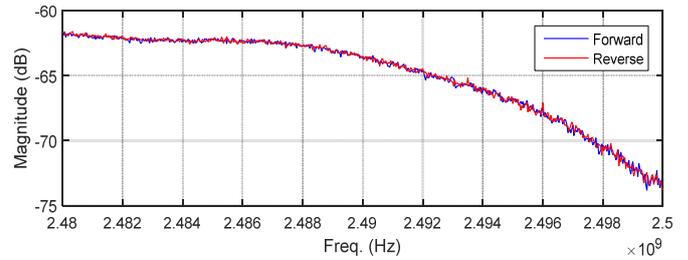
Fig. 5: Scenario 1 in Fig. 3, measurements, separation distance=7 m.

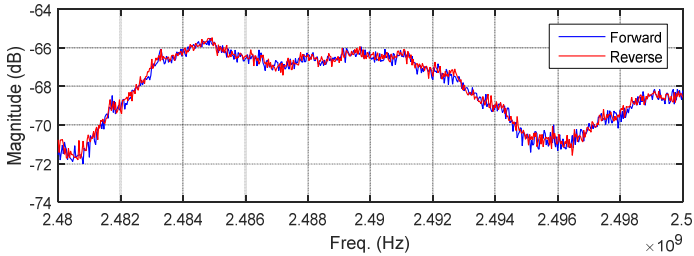
Fig. 6: Scenario 1 in Fig. 3, measurements, separation distance=10 m.

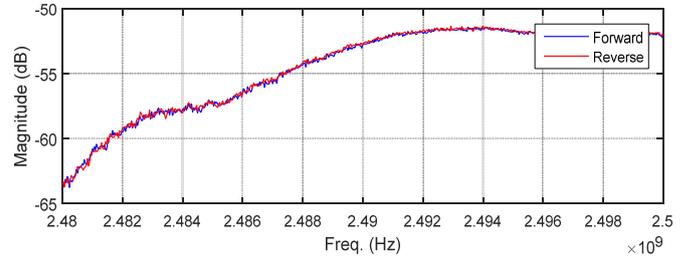
Fig. 7: Scenario 2 in Fig. 3, measurements, separation distance=5 m.

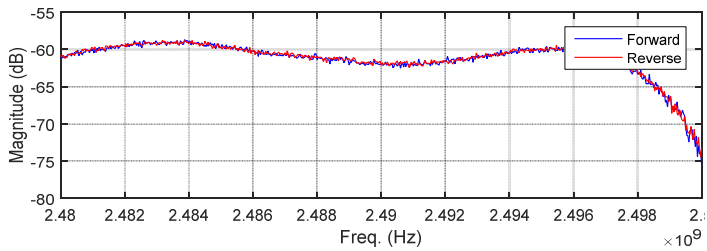
Fig. 8: Scenario 2 in Fig. 3, measurements, separation distance=9 m.

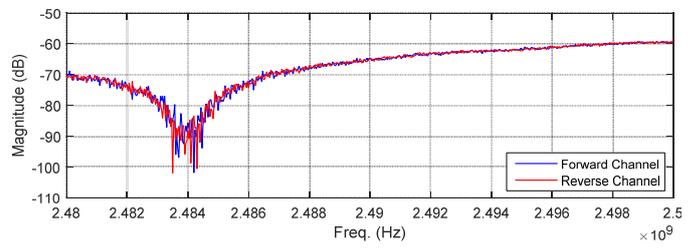
Fig. 9: Scenario 3 in Fig. 3, lab-to-room NLOS measurements (2m).

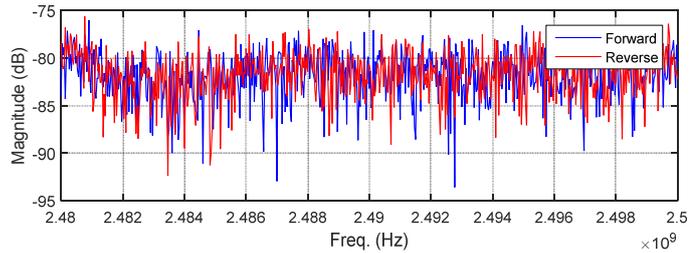
Fig. 10: Scenario 3 in Fig. 3, lab-to-room NLOS measurements (~4m).

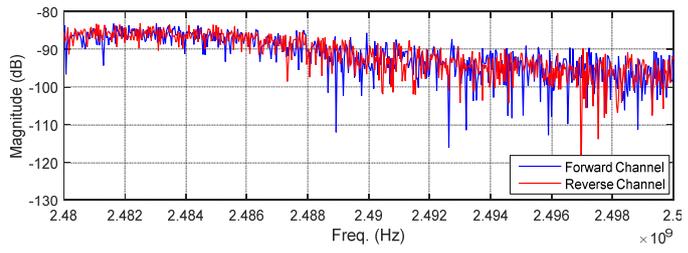
Fig. 11: Scenario 4 in Fig. 3, lab-to-corridor NLOS measurements.

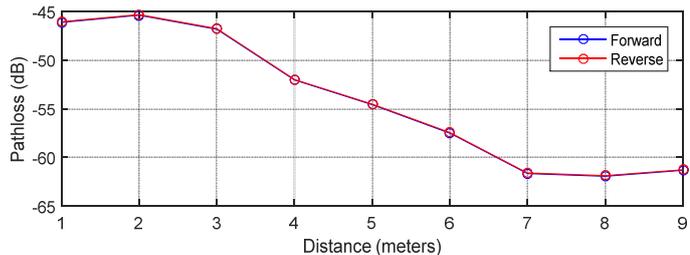
Fig. 12: Scenario 2 in Fig. 3, lab-139 LOS pathloss plots.

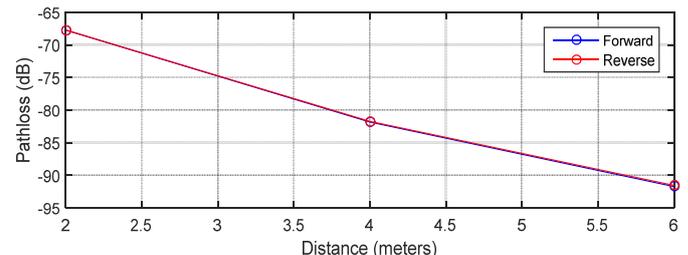
Fig. 13: Scenario 4 in Fig. 3, lab-to-corridor NLOS pathloss plots.

are not contributors to channel asymmetry.

For future research directions, the assumption of wireless channel asymmetry will be adopted to regenerate some studies (e.g. [24] – [25]) and the performance will be systematically compared. Moreover, further investigations are also planned for other wireless axioms. We also plan to analytically model the effect of the impairments on the wireless channels.

ACKNOWLEDGMENT

The authors are deeply thankful for Dr. Nayef Al-Sindi and Dr. James Aweya from the Etisalat British Telecom (BT) Innovation Center for their fruitful discussions and overall supervision of the work. The authors are also very thankful to Mrs. Sara Al-Shamsi for her assistance in collecting data.

## Resources

Measurements data and documentation as well as the used MATLAB codes in these experiments are available online for interested readers at [19] so that they can regenerate the plots, carry more statistical analysis or utilize them for possible other applications.